%% file: main.tex
\documentclass[a4paper,10pt]{article}
\usepackage[T1]{fontenc}
\usepackage{amsmath}
\usepackage{amssymb}
\usepackage{graphicx}
\usepackage{empheq}
\usepackage{lmodern}
\usepackage{float}
\usepackage{caption}
\usepackage{algorithm}
\usepackage{booktabs}
\usepackage{subcaption}
\usepackage{ulem}
\usepackage{multicol}
\usepackage[noend]{algpseudocode}
\usepackage{hyperref}
\usepackage{natbib}
\setcitestyle{authoryear,open={(},close={)}}
\usepackage{chngcntr}
\usepackage{placeins}

\usepackage[svgnames]{xcolor}

\usepackage{colortbl}
\usepackage{titlesec}

\usepackage{hyperref}
\makeatletter
\newcounter {subsubsubsection}[subsubsection]
\renewcommand\thesubsubsubsection{\thesubsubsection .\@alph\c@subsubsubsection}
\newcommand\subsubsubsection{\@startsection{subsubsubsection}{4}{\z@}%
                                     {-3.25ex\@plus -1ex \@minus -.2ex}%
                                     {1.5ex \@plus .2ex}%
                                     {\normalfont\normalsize\bfseries}}
\newcommand*\l@subsubsubsection{\@dottedtocline{3}{10.0em}{4.1em}}
\newcommand*{\subsubsubsectionmark}[1]{}
\makeatother

\hypersetup{
  colorlinks=true,
  citecolor=LimeGreen,
  linkcolor=Magenta,
  urlcolor=Blue
}
\def\be{\begin{eqnarray}}
\def\ee{\end{eqnarray}}

\DeclareMathOperator*{\argmax}{arg\,max}

\usepackage{geometry}
\usepackage{chngpage}
\usepackage{array}
\newcolumntype{R}[1]{>{\raggedleft\arraybackslash }b{#1}}
\newcolumntype{L}[1]{>{\raggedright\arraybackslash }b{#1}}
\newcolumntype{C}[1]{>{\centering\arraybackslash }b{#1}}
\newcolumntype{P}[1]{>{\centering\arraybackslash}p{#1}}

\usepackage{listings}

\def\1{{\bf 1}}

\usepackage{natbib}
\usepackage{bbm}

\input ./inputs/abrege
\input ./inputs/abrmath

\input ./inputs/alphabet
\input ./inputs/beginend

\begin{document}

\title{Deep combinatorial optimisation for optimal stopping time problems : application to swing options pricing.}
\author{Thomas DESCHATRE \footnote{EDF R\&D \& FiME, Laboratoire de Finance des March\'es de l'Energie} \thanks{thomas-t.deschatre@edf.fr}
\and
Joseph MIKAEL \footnote{EDF R\&D} \thanks{joseph.mikael@edf.fr}
}
\date{\today}
\maketitle

\abstract{
\textit{{A new method for stochastic control based on neural networks and using randomisation of discrete random variables is proposed and applied to optimal stopping time problems. The method models directly the policy and does not need the derivation of a dynamic programming principle nor a backward stochastic differential equation. Unlike continuous optimization where automatic differentiation is used directly, we propose a likelihood ratio method for gradient computation.
Numerical tests are done on the pricing of American and swing options. The proposed algorithm succeeds in pricing high dimensional American and swing options in a reasonable computation time, which is not possible with classical algorithms.
\\  \vspace{5mm} }}

{\bf Mathematics Subject Classification (2010).} 91G60, 60G40, 90C27, 97R40.

\vspace{5mm}
{\bf Keywords.} Optimal stopping, American option, Swing option, Combinatorial optimisation, Neural network, Artificial intelligence.}\\  \vspace{5mm} 

\noindent

\vspace{5mm}

\input{1_introduction}

\input{2_core}

\input{3_conclusion}
\FloatBarrier
\bigskip
{\bf Acknowledgements.} This research is supported by the department OSIRIS (Optimization, SImulation, RIsk and Statistics for energy markets) of EDF Lab which is gratefully acknowledged.
\bibliography{bibliography}

\end{document}

%% file: inputs/abrmath.tex
\newsavebox{\fminibox}
\newlength{\fminilength}
\newenvironment{fminipage}[1][\linewidth]
  {\setlength{\fminilength}{#1}
   \begin{lrbox}{\fminibox}\begin{minipage}{\fminilength}}
  {\end{minipage}\end{lrbox}\noindent\fbox{\usebox{\fminibox}}}

  \def\+{^\dagger}

\def\nequiv{\not\kern-.05em\equiv}
\def\egal{\kern-.5em=\kern-.5em}        
\def\propt{\kern-.2em\propto\kern-.2em} 

\def\argmax{\mathop{\mathrm{arg\,max}}} 
\def\intdouble{\int\kern-0.3em\int}
\def\inttriple{\int\kern-0.3em\int\kern-0.3em\int}

\def\rond#1{\overset{\kern-0.33em~_\circ}{#1}}
\def\rondit[#1]#2{\overset{\kern#1~_\circ}{#2}}


%% file: inputs/beginend.tex
\def\babs{\begin{abstract}}             \def\eabs{\end{abstract}}
\def\barr{\begin{array}}                \def\earr{\end{array}}
\def\bcc{\begin{center}}                \def\ecc{\end{center}}

\def\bdes{\begin{description}}          \def\edes{\end{description}}
\def\bdoc{\begin{document}}             \def\edoc{\end{document}}
\def\ben{\begin{enumerate}}             \def\een{\end{enumerate}}
\def\beqn{\begin{eqnarray}}             \def\eeqn{\end{eqnarray}}
\def\beqnl#1{\beqn\label{#1}}           \def\eeqnl#1{\label{#1}\eeqn}
\def\beqnx{\begin{eqnarray*}}           \def\eeqnx{\end{eqnarray*}}
\def\bseqn{\begin{subeqnarray}}         \def\eseqn{\end{subeqnarray}}
\def\beq#1\eeq{\begin{equation}#1\end{equation}}
\def\bal#1\eal{\begin{align}#1\end{align}}
\def\balx#1\ealx{\begin{align*}#1\end{align*}}
\def\beqx{$$}                           \def\eeqx{$$}
\def\bfig{\protect\begin{figure}}       \def\efig{\protect\end{figure}}
\def\bfigx{\protect\begin{figure*}}     \def\efigx{\protect\end{figure*}}
\def\bfigt{\protect\begin{figurette}}   \def\efigt{\protect\end{figurette}}
\def\bfl{\begin{flushleft}}             \def\efl{\end{flushleft}}
\def\bfr{\begin{flushright}}            \def\efr{\end{flushright}}
\def\bit{\begin{itemize}}               \def\eit{\end{itemize}}
\def\bmi{\begin{minipage}}              \def\emi{\end{minipage}}
\def\bfmi{\begin{fminipage}}            \def\efmi{\end{fminipage}}
\def\bpic{\begin{picture}}              \def\epic{\end{picture}}
\def\bqu{\begin{quote}}                 \def\equ{\end{quote}}
\def\bqun{\begin{quotation}}            \def\equn{\end{quotation}}
\def\bsl{\begin{slide}}                 \def\esl{\end{slide}}
\def\btabb{\begin{tabbing}}             \def\etabb{\end{tabbing}}
\def\btabl{\begin{table}}               \def\etabl{\end{table}}
\def\btablx{\begin{table*}}             \def\etablx{\end{table*}}
\def\btab{\begin{tabular}} 
\def\btabu{\begin{tabular}}             \def\etabu{\end{tabular}}
\def\btabx{\begin{tabular*}}            \def\etabx{\end{tabular*}}
\def\bbib{\begin{thebibliography}{999}} \def\ebib{\end{thebibliography}}
\def\bver{\begin{verbatim}}             \def\ever{\end{verbatim}}
\def\bca{\begin{cases}}                  \def\eca{\end{cases}}

%% file: 1_introduction.tex
\section{Introduction}
\label{sec:introduction}

\paragraph*{Motivation} Optimal stopping problems are particularly important for risk management as they are involved in the pricing of American options. American-style options are used not only by traditional asset managers but also by energy companies to hedge ``optimised assets'' by finding optimal decisions to optimise their P\&L and find their value. A common modelling of a power plant unit P\&L is done using swing options which are options allowing to exercise at most $l\geq 1$ times the option with possibly a constraint on the delay between two exercise dates (see \cite{carmona08} or \cite{warin12}).

\medskip
Formally, for $T > 0$, we are given a stochastic processes $(X_t)_{t \geq 0}$ defined on a probability space \\
$\left(\Omega, \mathcal{F}, \mathbb{F} = (\mathcal{F}_t)_{t \geq 0}, \mathbb{P}\right)$ and we search for an increasing sequence of $\mathbb{F}$ stopping times $\tau = (\tau_1, \tau_2, \ldots, \tau_l)$ maximizing the expectation of the objective function 
\[\mathbb{E}_{\mathbb{P}}\left( \sum_{i=1}^l f(\tau_i, X_{\tau_i}){\bf 1}_{\tau_i \leq T} \right).\]

Numerical methods to solve the optimal stopping problem when $l = 1, f(x,t) = e^{-rt}g(x)$ and $X$ is Markovian include:
\begin{itemize}
    \item Dynamic programming equation: the option price $P_0$ is computed using the following backward discrete scheme over a grid $t_0 = 0 < t_1 < \ldots < t_N = T$:
    \begin{equation} \label{eq:dynprog}
    \begin{split}
    &P_{t_N} = g(X_T),\\
    &P_{t_{i}} = \max(g(X_{t_i}), e^{-r(t_{i+1}-t_i) }\mathbb{E}_{\mathbb{P}}(P_{t_{i+1}} | \mathcal{F}_{t_i})), \; i=0,\ldots,N-1.
    \end{split}
    \end{equation}
    One then needs to perform regression to compute the conditional expectations, see \cite{longstaff2001valuing} or \cite{bouchard2012monte}. 
    \item Partial differential equation (PDE): a variational inequality  derived from the Hamilton Jacobi Bellman equation is given by 
    \[\min(- (\partial_t + \mathcal{L})v + rv, v-g) = 0, \; v(x, T) = g(x)\]
    where $\mathcal{L}$ is the infinitesimal generator of $X$ \cite[Chapter 8, Section 3.3]{shreve04}. A numerical scheme can be applied to solve this PDE and find the option  value.
    \item Reflected Backward Stochastic Differential Equation (BSDE): the value of the American option is the solution of the reflected BSDE \cite{el97}:
    \[Y_t = g(X_T) - r \int_t^T Y_s ds - \int_t^T Z_s dW_s + K_T - K_t,\]
    \[Y_t \geq g(X_t), \; 0 \leq t \leq T,\]
    \[\int_0^T (Y_t - g(X_t))dK_t = 0.\]
    \cite{bouchard08} provides a numerical scheme to solve these equations. 
    \item Policy search: the decision rule or exercise region is parametrized by a vector and the parameters are usually optimised by Monte Carlo methods as in reinforcement learning \cite[Chapter 8, Section 2]{glasserman13}; \cite{andersen1999, garcia2003}. The algorithm proposed in this paper is strongly related to this class of method.
\end{itemize}
These approaches generalise well for $l \geq 1$, see \cite{carmona08} for dynamic programming principle or \cite{bernhart12} for the BSDE method. The non linear case where $f$ is of the form $\phi(\sum_{i=1}^l e^{-r \tau_i} g(X_{\tau_i}) {\bf 1}_{\tau_i \leq T})$ is studied by \cite{trabelsi13}. We refer to \cite[Chapter 8]{glasserman13} for more exhaustive details on numerical methods for American option pricing. All these algorithms suffer from the curse of dimensionality: the number of underlying is hardly above 5. However energy companies portfolio may trade derivatives involving more that 4 commodities at one time (e.g. swing options indexed on C02, natural gas, electricity, volume, fuel) and traditional numerical methods hardly provide good solutions in a reasonable computing time. 

\medskip

Recently, neural network-based approaches have shown good results regarding stochastic control problems and PDE numerical resolution in high dimension, see \cite{han2017overcoming,sirignano2018,Chan-Wai-Nam2019}. In the following, one describes literature related to optimal stopping time problems using neural networks. \cite{kohler2010, becker2020pricing} use neural networks for regression in the dynamic programming equation \eqref{eq:dynprog}. \cite{hure18, bachouch2018, becker19a} also use the dynamic programming equation \eqref{eq:dynprog} but neural networks are used to parameterize the optimal policy. Weights and bias of the neural network(s) minimise at each time step the right hand side of the dynamic programming equation \eqref{eq:dynprog}, going backward. The optimal decision consists in a continuous variable (instead of a discrete one) taking value in $\left(0,1\right)$ modeled by a neural network. 
\cite{han17, weinan2017deep,hure19} use neural networks to solve BSDE's. In \cite{hure19}, the neural networks parameterizing the solution and eventually its gradient minimise the $L^2$ loss between the left hand-side and the right hand side of the Euler discretisation of the BSDE, going backward from the terminal value. \cite{bachouch2018} and \cite{hure19} need to maximise one criteria by time step. The approaches of \cite{han17} and \cite{weinan2017deep} are quite different: the neural network allows the parameterisation of the initial value of the BSDE and the gradient at each time step, and it minimises the distance between the terminal value obtained by the neural network and the terminal value of the BSDE, going forward. American put options prices are computed in \cite{hure19} up to dimension 40 with 160 time steps. Neural networks approaches have also been used in the context of swing options pricing in gas market in \cite{barrera06}. The definition of swing options slightly differs from ours as it considers a continuous control: the option owner buys a certain amount of gas between a minimum and a maximum quantity. It is however related to our problem as in continuous time, this option is bang-bang: it is optimal to exercise at the minimum or the maximum level at each date, that is choosing between two actions. \cite{barrera06} directly models the policy by a neural network and optimises the objective function as in \cite{fecamp2019risk, buehler2019}. 



\medskip
Contrarily to \cite{kohler2010, hure18, bachouch2018, han17, weinan2017deep, hure19, becker2020pricing}, the goal of this paper is to propose a reinforcement learning algorithm to solve optimal multi-exercise (rather than one single) stopping time problems with constraints on exercise times that does not need to derive a dynamic programming equation nor to find an equivalent BSDE of the problem. The only information needed is the dynamic of the state process $X$ and the objective function. This kind of algorithm is called policy gradient and is well known in the area of reinforcement learning, see \cite{sutton00} for instance. Although continuous control approximation with reinforcement learning shows good results, see \cite{fecamp2019risk, buehler2019} for European-style option hedging, the case of optimal stopping times is more difficult as it involves controls taking values in a discrete set of actions. The problem is similar to a combinatorial optimisation one: at each time step, an action belonging to a finite set needs to be taken. One way to solve this problem is to perform a relaxation assuming that the control belongs to a continuous space. For instance, if one needs to price an American option, a decision represented by a value in $\{0,1\} $ and consisting in exercising or not must be taken. Relaxing the problem consists in searching for solutions in $\left[0,1\right]$: this relaxation has successfully been applied to a Bermudan option pricing in a high dimensional setting (up to 1000) in \cite{becker19b}. These methods apply well for American-style option pricing but seem to be not flexible enough to be extended to swing options pricing.


\medskip
\paragraph*{Main results} Our approach follows the spirit of \cite{fecamp2019risk} and \cite{becker19b}: one directly parameterises the optimal policy by a neural network and maximises the objective function moving forward. 
We propose an algorithm using reinforcement learning in order to solve optimal stopping times problem seen as an combinatorial optimisation problem. Note that solving combinatorial optimisation problems with neural networks have been considered in \cite{bello2016neural} in a deterministic framework without dynamics on the state process. The stochastic optimization framework considered in this paper is described in Section \ref{sec:optimalstopping}.

Neural network hardly handles integer outputs which is the main difficulty of the problem addressed in this paper. To encompass this problem, the first step of the algorithm consists in randomizing the optimization variables (that is executing the option or not) and modelling their law by a neural network. The second step consists in computing the gradient of the objective function. It can not be computed as usual by automatic differentiation as the neural network does not output the optimization variables but their law. The use of likelihood ratio method allows to rewrite the gradient as a function of the neural network output gradient that can be computed with automatic differentiation. The algorithm is given in Section \ref{sec:optimalstoppingalgorithm}. Compared to the papers referenced above our approach allows to solve stopping time problems without any knowledge of the dynamic programming equation or of an equivalent BSDE. Furthermore, it presents many advantages as it
\begin{itemize}
    \item can solve multiple optimal stopping time problems;
    \item allows to add in a flexible way any constraint on the stopping times;
    \item can then be associated with the one of \cite{fecamp2019risk} considering continuous actions in order to solve stochastic impulse control problems, combining discrete and continuous controls (see \cite[Chapter 6]{oksendal2005applied} for more information on impulse control problems).
\end{itemize}  
Let us also notice that our method does not take advantage of the linearity (possible inversion between the sum and the expectation) of the considered optimal stopping problems contrarily to \cite{becker19b} and can be applied to non linear problem, making it suitable for impulse control problems. The theoretical convergence study of our algorithm is out of the scope of this paper.

 Numerical tests covering Bermudan and swing options are proposed in Section \ref{sec:numericalresultsoptimalstopping} and show good results in the pricing of 10 underlyings Bermudan option and also on 5 underlyings swing options having up to $l = 6$ exercise dates. However, our algorithm gives suboptimal results on one of the considered case.

%% file: 2_core.tex
\section{Optimal stopping}
\label{sec:optimalstopping}
\subsection{Continuous time modelling}
\label{sec:continuoustimemodelling}
We are given a financial market operating in continuous time. Let $\left(\Omega, \mathbb{F}= (\mathcal{F}_t)_{t \geq 0}, \mathcal{F}, \mathbb{P}\right)$ a filtered probability space and $W$ a d-dimensional $\mathbb{F}$-Brownian motion. One assumes that $\mathbb{F}$ satisfies the usual conditions of right continuity and completeness. Let $T > 0$ a finite horizon time and $X = (X_1, X_2, \ldots, X_d)$ be the  unique strong solution of the Stochastic Differential Equation (SDE):
\begin{equation} \label{sde}
X_t = X_0 + \int_0^t \mu(s, X_s) ds + \int_0^t \sigma(s, X_s)dW_s, \; t \in \left[0,T\right],
\end{equation}
with $\mu : \left[0,T \right] \times \mathbb{R}^d \mapsto \mathbb{R}^d$ and $\sigma: \left[0,T \right] \times \mathbb{R}^d \mapsto \mathbb{R}^{d \times d}$ two measurable functions verifying $|\mu(t, x) - \mu(t, y)| + \|\sigma(t, x) - \sigma(t, y)\| \leq K_1 |x -y|$ and $|\mu(t,x)| + \|\sigma(t,x)\| \leq K_2(1 + |x|)$ for $x,y\in\mathbb{R}^d$ and $t\in[0,T]$ ($|\cdot|$ denotes the Euclidian distance in $\mathbb{R}^d$ and for a matrix $A \in \mathbb{R}^{d \times d}, \|A\| = \sqrt{tr(A A^{\top})}$) and $K_1, K_2\in \mathbb{R}$. Using the notations of \cite{carmona08} and with $X$ as defined in \eqref{sde} for $t \in \left[0,T\right]$ and $X_t = X_T$ for $t \geq T$, an optimal stopping time problem consists in solving the problem 
\begin{equation} \label{eq:optimalstoppingcontinuous}\underset{\tau \in \mathcal{S}^{l}}{\sup} \mathbb{E}_{\mathbb{P}}\left( \sum_{i=1}^l f(\tau_i, X_{\tau_i}) {\bf 1}_{\tau_i \leq T}\right)
\end{equation}
where $\mathcal{S}^l$ is the collection of all vectors of increasing stopping times $\tau = (\tau_1, \ldots, \tau_l)$ such that for all $i = 2, \;\ldots,l$,  $\tau_i - \tau_{i-1} \geq \gamma$ a.s. on the set of events $\{\tau_{i-1} \leq T\}$ and where $f: \left[0,T\right] \times \mathbb{R}^d \mapsto \mathbb{R}$ is a measurable function. $l \in \mathbb{N}^* = \mathbb{N} \setminus \{0\}$ corresponds to the number of possible exercises and $\gamma \geq 0$ to the minimum delay between two exercise dates. 
One wants to find the optimal value \eqref{eq:optimalstoppingcontinuous} but also the optimal policy
\begin{equation} \label{eq:optimalcontrol}
\tau^* \in \underset{\tau \in \mathcal{S}^{l}}{\text{argmax }} \mathbb{E}_{\mathbb{P}}\left(\sum_{i=1}^l f(\tau_i, X_{\tau_i}) {\bf 1}_{\tau_i \leq T}\right).
\end{equation}

\subsection{Discrete time modelling}
\label{sec:discretetimemodelling}
In practice, one only considers optimal stopping on a discrete time grid (for instance, the valuation of a Bermudan option is used as a proxy of the American option). Let us consider $N + 1$ exercise dates belonging to a discrete set $\mathcal{D}_N = \{t_0 = 0 < t_1 < \ldots < t_N = T\}$, $N \in \mathbb{N}^*$. The problem consists in finding 
\begin{equation} 
  \underset{\tau \in \mathcal{S}^l_N}{\sup \;} \mathbb{E}_{\mathbb{P}}\left( \sum_{i=1}^l f(\tau_i, X_{\tau_i}) {\bf 1 }_{\tau_i \leq T}\right) \label{eq:base_problem}
\end{equation}
where $S^l_N$ is the set of stopping times belonging to $S^l$ such that $\tau_i \in \mathcal{D}_N$ on $\{\tau_i \leq T\}$, for $i=1,\ldots,l$.
This discretisation is needed for our algorithm as it is needed in classical methods such as \cite{longstaff2001valuing}. Problem \eqref{eq:base_problem} is equivalent to the following:
\begin{eqnarray}
\underset{Y}{\sup \;} \mathbb{E}_{\mathbb{P}}\left( \sum_{i=0}^N Y_i f(t_i, X_{t_i})\right) \label{eq:discrete_base_problem}
\end{eqnarray}
where $(Y_i)_{i=0,\ldots,N}$ is a sequence of  $(\mathcal{F}_{t_i})_{i=0,\ldots,N}$-measurable random variables taking values in $\{0,1\}$ such that 
\begin{equation} \label{eq:constraintsum} \sum_{i=0}^N Y_i \leq l
\end{equation}
and 
\begin{equation}
    \label{eq:constraintdelay} D_j \geq \gamma, \; j=0,\ldots,N,
\end{equation}
with 
\begin{equation}\label{eq:delay}
    D_j = \gamma + t_j - \sum_{i=0}^{j-1} Y_i D_i
\end{equation}
the delay at $t_j$ from the last exercise assuming that we can exercise at time 0 (hence the $\gamma$ term in \eqref{eq:delay} allowing to start at time 0 with a delay $\gamma$) and with the convention $\sum_{i=0}^{-1} \cdot = 0$.
Given a solution $(Y_i^*)_{i=0,\ldots,N}$ of Problem \eqref{eq:discrete_base_problem}, a proxy for the optimal control \eqref{eq:optimalcontrol} is given by 
\[\tau^*_k = t_{m(k)}{\bf 1}_{k \leq m} + \infty {\bf 1}_{k > m}, \; k \in \{1, \ldots, l\},\]
on the event $\{\sum_{i=0}^N Y_i = m\}$ with $m \leq l$ and $m(k) = \min \{j \in \{0,\ldots,N\} | \sum_{i=0}^j Y^*_i \geq k\}$. 

\section{Algorithm description}
\label{sec:optimalstoppingalgorithm}

\subsection{Neural network parametrization}

As the $Y_i$'s are discrete we cannot assume that they are the output of a neural network which weights are optimised by applying a stochastic gradient descent (SGD). To overcome this difficulty, one can randomize $Y$ and consider that at each time step $t_j, j \in \{1,\ldots, N\}$, the discrete variable $Y_j$ is a Bernoulli distributed random variable conditionally on $\mathcal{F}_{t_{j}} \cup \sigma(Y_i, i \leq j-1)$ (and $\mathcal{F}_{t_{0}}$ if $j=0$). The Bernoulli distribution parameter depends on the state variable of our control problem. This state variable, denoted $S_{t_j}$, is
\begin{itemize}
\item $X_{t_j}$ in the case of a Bermudan option, that is with only one execution date, 
\item $(X^1_{t_j}, \ldots, X^d_{t_j}, \sum_{i=0}^{j-1} Y_i,D_j)$  when there are constraints \eqref{eq:constraintsum} and \eqref{eq:constraintdelay}.
\end{itemize}
Of course, one can adapt the state depending on the constraints (for instance if there is no delay constraint \eqref{eq:constraintdelay}, the state is $(X^1_{t_j}, \ldots, X^d_{t_j}, \sum_{i=0}^{j-1} Y_i)$). In a non Markovian framework, one could consider that the probability for $Y_j$ to be equal to 1 is a function of all the values of $X_{t_i}$ for $i \leq j$ and $Y_i$ for $i \leq j-1$. In this case, one could use a Recurrent Neural Network to parameterise this function but we do not consider this case here. 
The Bernoulli distribution parameter, which is $\mathbb{P}(Y_{j} = 1 | S_{t_j})$ is parameterised by a neural network $\mathbb{NN}$ defined on $\left[0,T\right] \times \mathbb{S} \times \Theta$ and taking values in $\mathbb{R}$ where $\mathbb{S}$ is the state space and $\Theta$ represents the sets in which the biases and weights of the neural network lie. The neural network architecture is described in Section \ref{sec:architectureoptimal}. The parametrization is then the following:
\begin{equation}
\label{eq:paramNN}
\mathbb{P}(Y_j = 1 | S_{t_j}) = expit\left( C \tanh\left(\mathbb{NN}(t_j, S_{t_j},\theta)\right)\right) c(S_{t_j}), \; j = 0,\ldots,N,
\end{equation}
with $expit: \mathbb{R} \mapsto \left(0,1\right)$ and $expit(x) = \frac{1}{1 + e^{-x}}$ for $x \in \mathbb{R}$ and $c(S_{t_j})$ is equal to 0 if the constraints are saturated and 1 otherwise. Typically, if there are no constraints, $c$ is always equal to 1, and if there are constraints \eqref{eq:constraintsum} and  \eqref{eq:constraintdelay},
\[
c(S_{t_j}) = {\bf 1}_{\sum_{i=0}^{j-1} Y_i < l} {\bf 1}_{D_j \geq \gamma}. 
\]
Note that the methodology can be extended to any constraint on the policy. $C  \tanh(\mathbb{NN}(x,\theta))$ outputs the $logit$ (the inverse function of $expit$) of $\mathbb{P}(Y_j = 1 | S_{t_j})$ (when constraints are not saturated). The function $tanh$ is not necessary and one could only consider $\mathbb{NN}$ to parameterise the logit of the probability. To reduce the values taken by the $logit$, we bound the output of the neural using $tanh$ and choose $C$ such that $expit(-C) \approx 0$ and $expit(C) \approx 1$ ; $C$ is given in Section \ref{sec:optimalstoppinghyperparameters}. 

From now on, $\mathbb{P}$ is replaced by $\mathbb{P}_{\theta}$ to indicate the dependence of the law of $Y$ with $\theta$. At this step, we still cannot train our neural networks by applying a stochastic gradient descent because of the $Y$'s randomization. 

\subsection{Optimization}

To approximate a solution to \eqref{eq:discrete_base_problem} we search for
\[
    \theta^{*}  \in \argmax_{\theta\in\Theta } \mathbb{E}_{\mathbb{P}_{\theta}}\left( \sum_{i=0}^N Y_i f(t_i, X_{t_i})\right).
\]
Classical neural network parameter optimization consists first in evaluating the objective function $$\mathbb{E}_{\mathbb{P}_{\theta}}\left( \sum_{i=0}^N Y_i f(t_i, X_{t_i})\right)$$ using Monte Carlo method and replacing it by 
\[
\frac{1}{N_{batch}} \sum_{m=1}^{N_{batch}} \sum_{i=0}^N Y^m_i f(t_i, X^m_{t_i})
\]
with $N_{batch} \in \mathbb{N}^{*}$ and where $X^m$ and $Y^m$ correspond to one realization of $(X, Y)$ simulated according to \eqref{sde} for $X$ and to $\mathbb{P}_{\theta}$ for $Y$. Secondly, a gradient descent is done to update the parameter $\theta$ using gradient of the objective function which is computed using backpropagation. However in our case, it is not possible to directly use backpropagation: $Y$ is not a function of $\theta$, but a discrete variable with law depending on $\theta$.

\medskip
To encompass this problem, we use a likelihood ratio method. Let us consider a random variable $Z : \Omega \mapsto E$ with probability measure $\mathbb{P}_a$, $a \in \mathbb{R}^d$, absolutely continuous with respect to a measure $\mathbb{P}$. Let $l_a(x) = \frac{d\mathbb{P}_a}{d\mathbb{P}}(x)$ be the likelihood function. We have, under some integrability conditions, $\nabla_a \mathbb{E}_{\mathbb{P}_a}(Z) = \mathbb{E}_{\mathbb{P}_a}(Z \nabla_a \log(l_a(Z)))$. 
Using this method and iterative conditioning for probability computation, we find that the gradient of $\mathbb{E}_{\mathbb{P}_{\theta}}\left( \sum_{j=0}^N Y_j f(t_j, X_{t_j})\right)$ is given by:
\begin{equation}
\begin{split}
 \label{eq:derive1} &\nabla_{\theta} \mathbb{E}_{\mathbb{P}_{\theta}}\left( \sum_{j=0}^N Y_j f(t_j, X_{t_j})\right) =\\
 &\mathbb{E}_{\mathbb{P}_{\theta}}\left( \sum_{j=0}^N Y_j f(t_j, S_{t_j})   \sum_{i=0}^N Y_i\nabla_{\theta}\log \left(\mathbb{P}_{\theta}\left(Y_i=1 | S_{t_i}\right)\right) + (1 - Y_i)\nabla_{\theta}\log \left(1 - \mathbb{P}_{\theta}\left(Y_i=1 | S_{t_i}\right)\right)\right).
 \end{split}
 \end{equation}
$\mathbb{P}_{\theta}(Y_i = 1 | S_{t_i})$ which is defined in Equation \eqref{eq:paramNN} is a continuous function of the neural network: the gradients appearing in Equation \eqref{eq:derive1} can be easily computed using backpropagation. Let us notice that the method does not take any advantage of the fact that we can exchange the sum and the expectation in \eqref{eq:discrete_base_problem}. It is then suitable for optimization problems of the form 
\[
\underset{Y}{\sup \;} \mathbb{E}_{\mathbb{P}}\left(g(Y_1, t_1, X_{t_1}, \ldots, Y_N, t_N, X_{t_N})\right) 
\]
and could be combined with forward neural network continuous optimization algorithms such as those of \cite{fecamp2019risk,buehler2019} to solve impulse control problems.

\medskip
Every $\Delta_{test}$ steps, the objective value is computed over the testing set. The parameters kept at the end are the ones minimising those evaluations. The objective function is finally evaluated on a validation set. While on the training phase actions are sampled from the outputted probability on the training set, they are chosen equal to 1 if the probability is greater than 0.5 and 0 otherwise on the test and validation sets. The algorithm is described in \ref{algo:algoGlobal} with hyperparameters given in Section \ref{sec:optimalstoppinghyperparameters}.

\begin{algorithm}[ht]
\begin{algorithmic}[1]
 \State $\alpha$ : Learning rate
 \State $\beta_1$, $\beta_2\in[0,1]:$ Exponential decay rates for the moment estimates,
 \State $N_{iter}:$ number of iterations
 \State $N_{batch}:$ number of simulations at each gradient descent iteration (batch size) 
\State $\theta_0$ randomly chosen 
\State $m_0 \leftarrow 0 $
\State $v_0 \leftarrow 0$ 
\For{$i_{iter}=1\ldots N_{Iter}$}
\For{$u=0 \ldots N$}
\State $X_u \leftarrow N_{batch}$ samples simulations of $X_{t_u}$
\State $S_u \leftarrow $ state value
\State \textcolor{blue}{
$p_{\theta_{i_{iter}-1},u} \leftarrow expit\left(C \tanh\left(\mathbb{NN}(t_u, S_u,\theta_{i_{iter}-1})\right)\right)c(S_u)$}
\State \textcolor{blue}{$Y_u \leftarrow N_{batch}$ of a Bernouilli r.v. with parameter $p_{\theta_{i_{iter}-1},u}$}
\EndFor
\State {\color{blue} 
$
g_{i_{iter}} \leftarrow  \frac{1}{N_{batch}} \sum_{n=1}^{N_{batch}} \left(\sum_{u=0}^N Y^n_u f(t_u, X^n_u) \right) 
\left(\sum_{u=0}^{N} Y^n_u \nabla_\theta \log\left(p_{\theta_{i_{iter}-1},u}\right) + (1-Y^n_u)\log\left(1-p_{\theta_{i_{iter}-1},u}\right)\right)$}
\State  $m_{i_{iter}} \leftarrow m_{{i_{iter-1}}} + (1-\beta_1) g_{i_{iter}}$ (update biased first moment estimate) 
\State $v_{i_{iter}} \leftarrow \beta_2 v_{{i_{iter-1}}} + (1-\beta_2) g_{i_{iter}} ^2$ (update biased second raw moment estimate)
\State $\hat{m}_{i_{iter}}  \leftarrow \frac{m_{i_{iter}}}{1-(\beta_1)^{i_{iter}}}$ (computes bias-corrected first moment estimate ) 
\State $\hat{v}_{i_{iter}}  \leftarrow \frac{v_{i_{iter}}}{1-(\beta_2)^{i_{iter}}}$ (computes bias-corrected second raw moment estimate)
\State $\theta_{i_{iter}} \leftarrow \theta_{i_{iter}-1} - \alpha \hat{m}_{i_{iter}} /(\sqrt{\hat{v}_{i_{iter}}}+\epsilon )$ (update parameters)
\EndFor
\end{algorithmic}
\caption{Algorithm for optimal stopping. The lines in blue are the main difference with classical backpropagation.}
\label{algo:algoGlobal}
\end{algorithm}

\subsection{Neural network architecture}
\label{sec:architectureoptimal}
The neural network architecture is inspired by \cite{Chan-Wai-Nam2019} and consists in one single feed forward neural network which features are the time step $t_i$ and the current state realisation $S_{t_i}$. Let $t \in \left[0,T\right]$ and $x = (x_1, \ldots, x_r)^{\top} \in \mathbb{S}$ that we assume to be included in $\mathbb{R}^r$. The neural network is defined as follow 
\[\mathbb{NN}(t, x, \theta) = A_{L+1} \circ \rho \circ A_{L} \circ \rho \ldots \circ A_{1} ((t, x_1, \ldots, x_r)^{\top})\]
where $A_l(y) = W_l y + b_l$ for $l =1,\ldots,L+1$, $W_1 \in \mathbb{R}^{(r+1) \times m}$, $W_l \in \mathbb{R}^{m \times m}$ for $l=2,\ldots,L$, $W_{L+1} \in \mathbb{R}^{m \times 1}$, $b_l \in \mathbb{R}^m$ for $l=1,\ldots,L$,  $b_l \in \mathbb{R}^{m}$ for $l=1,\ldots,L$ and $b_{L+1} \in \mathbb{R}$. $L$ corresponds to the number of layers and $m$ to the number of neurons per layer (that we assume to be the same for every layer). The $(W_l)_{l=1,\ldots,L+1}$ correspond to the weights and $(b_l)_{l=1,\ldots,L+1}$ to the biases. The function $\rho$ is the activation function and is chosen as the ReLu function, that is $\rho(x) = \max(0, x)$. $\theta$ is then equal to $(W_1, \ldots, W_{L+1}, b_1, \ldots, b_{L+1})$.


\subsection{Hyper parameters}
\label{sec:optimalstoppinghyperparameters}
\begin{itemize}
    \item The {\bf batch size} $N_{batch}$ as the {\bf number of iterations} $N_{iter}$ depend on the use case and are specified at a later stage. The training set size is then equal to $N_{iter} \times N_{batch}$. As the likelihood ratio estimator of the gradient has high variance, choosing a large batch size (>1000) allows for a better estimation. The drawback is that it tends to slow down the algorithm. To reduce the variance, one could also use a baseline function as in \cite{zhao11}.
    \item The {\bf test set size} is chosen equal to 500,000 and the {\bf validation set size} to 4,096,000 (500,000 and 4,096,000 are chosen high to have very accurate optimisation). The test set is evaluated every $\Delta_{test} = 100$ steps.
    \item The \textbf{number of layers} is chosen equal to 3. The \textbf{number of neurons} per layer is constant (but can vary from a case to another). 
    \item The \textbf{learning rate} ($\alpha$ in Algorithm \ref{algo:algoGlobal}) is chosen equal to 0.001. 
    \item As in \cite{Chan-Wai-Nam2019}, in our case, \textbf{regularisation} which is classically used to avoid overfitting is not relevant and we won't use it as our data is not redundant and thus the network does not experience overfitting.
    \item Since we use the same network at each time step, we use a mean-variance {\bf normalisation} over all the time steps to center all the inputs $(t_i, X_{t_i})$ for all $t_i$'s with the same coefficients. The scaling and recentering coefficients are estimated on 100,000 pre-simulated data that is just used to this end. The mean and the standard deviation are first computed for every time step over the simulations then averaged over the time steps.
    \item We use Xavier \textbf{initialisation} \cite{glorot2010understanding} for the weights and a zero initialisation for the biases.
    \item The parameter $C$ that bounds the input of the $expit$ function is chosen such that $expit(-C) \approx 0$ and $expit(C) \approx 1$. We choose $C = 10$. 
    \item The library used is \cite{2015tensorflow} and the algorithm runs on a laptop with 8 cores of 2,50 GHz, a RAM memory of 15,6 Go and without GPU acceleration.
\end{itemize}


\section{Numerical results}
\label{sec:numericalresultsoptimalstopping}

In this section Algorithm \ref{algo:algoGlobal} is applied to the valuation of Bermudan and swing options. The function $f(s,x)$ is of the form $e^{-rs}g(x)$ where $g$ is the payoff of the option and $r \geq 0$ is the risk free rate. We place ourselves in the Black-Scholes framework: $\mu(s, x) =  (r - \delta) x$, with $\delta \geq 0$ corresponding to the dividend rate and $\sigma(s, x) = diag(x) \sqrt{\Sigma}^{\top}$ with $\Sigma$ a positive definite matrix. We choose to work with a regular time grid $t_i = \frac{i}{N}$ for $i=0,\ldots,N$. The probability measure corresponds to the risk neutral probability and finding the value of the option consists in solving Problem \eqref{eq:discrete_base_problem}.

\subsection{Bermudan options}

\label{sec:american}
In this section, we assume that $l = 1$ (only one exercise) and we consider different options to price. 

\paragraph*{Put option}with $d = 1$, payoff $g(x) = (K-x)^+$, $K = 1$, $S_0 = 1$, $r = 0.05$, $\delta = 0$, $\sqrt{\Sigma} = 0.2$, $N = 10$, $T =1$. We consider a batch size equal to $N_{batch} = 5,000$, a neural network with a depth of 3 layers having 10 neurons each and $N_{iter} = 5,000$ iterations.

\paragraph*{Max-call option}with $d \in \{2, 10\}$, payoff $g(x) = (\max((x_i)_{i=1,\ldots,d}) - K)^+$, $K = 100$, $S_0^i = 100$, $i=1,\ldots,d$, $r = 0.05$, $\delta = 0.1$, $\sqrt{\Sigma} = 0.2  I_{d}$ ($I_d$ is the identity matrix with size $d \times d$), $N = 9$, $T = 3$. We consider a batch size equal to $N_{batch} = 5,000$ for $d = 2$ and $N_{batch} = 12,000$ for $d = 10$, a neural network with 3 layers of size 30 for $d = 2$ and 70 for $d = 10$ and $N_{iter} = 10,000$ iterations.

\paragraph*{Strangle spread option} with $d = 5$, payoff $g(x) = -(K_1 - \frac{1}{5} \sum_{i=1}^5 x_i)^+ + (K_2 - \frac{1}{5} \sum_{i=1}^5 x_i) + (\frac{1}{5} \sum_{i=1}^5 x_i - K_3)^+ - (\frac{1}{5} \sum_{i=1}^5 x_i - K_4)^+$, $K_1 = 75$, $K_2 = 90$, $K_3 = 110$, $K_4 = 125$, $S_0^i = 100, \; i=1,\ldots,5$, $r = 0.05$, $\delta = 0$, $$\sqrt{\Sigma} = \begin{pmatrix} 
0.3024 & 0.1354 & 0.0722 & 0.1367 & 0.1641 \\ 
0.1354 & 0.2270 & 0.0613 & 0.1264 & 0.1610 \\
0.0722 & 0.0613 & 0.0717 & 0.0884 & 0.0699 \\
0.1367 & 0.1264 & 0.0884 & 0.2937 & 0.1394 \\
0.1641 & 0.1610 & 0.0699 & 0.1394 & 0.2535
\end{pmatrix},$$
$N = 48$, $T = 1$. We consider a batch size equal to $N_{batch} = 8,000$, a neural networks with 3 layers size 60 and $N_{iter} = 10,000$ iterations. 

\begin{figure}[!ht]
    \centering
    \begin{subfigure}[b]{0.45\textwidth}
        \centering
        \includegraphics[width=\textwidth]{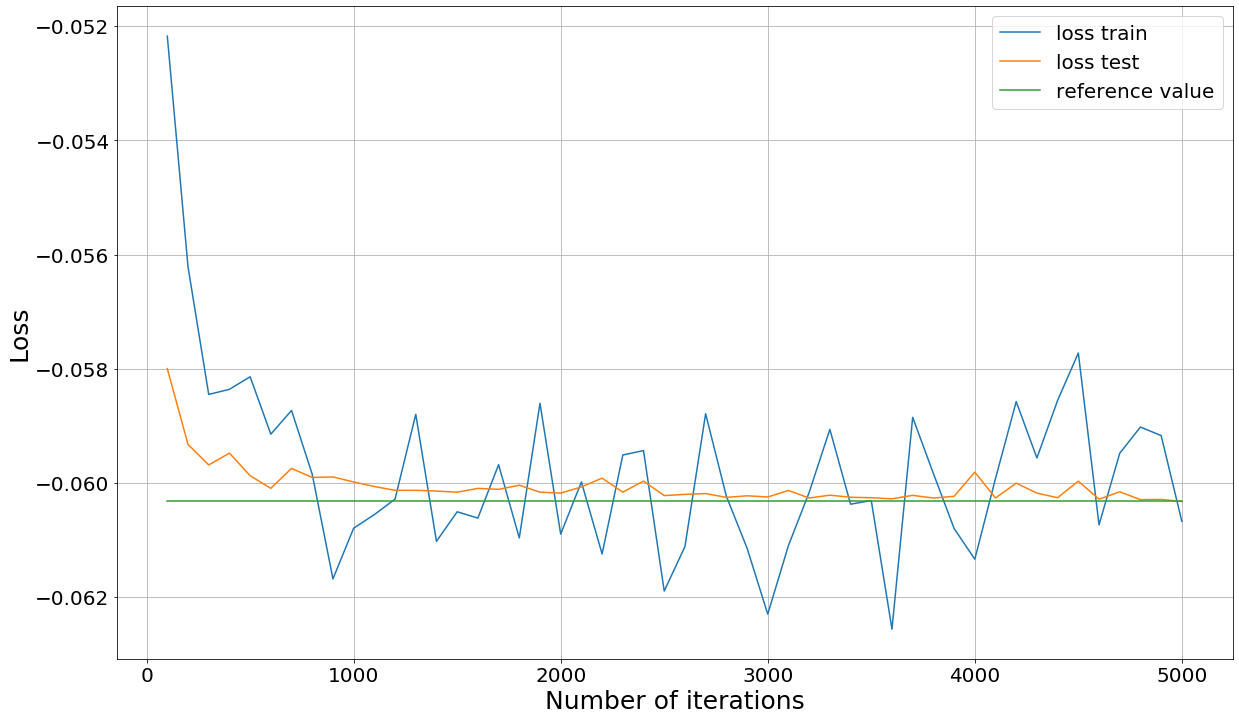}
        \caption{\it Put.}
    \end{subfigure}
    \begin{subfigure}[b]{0.45\textwidth}
        \centering
        \includegraphics[width=\textwidth]{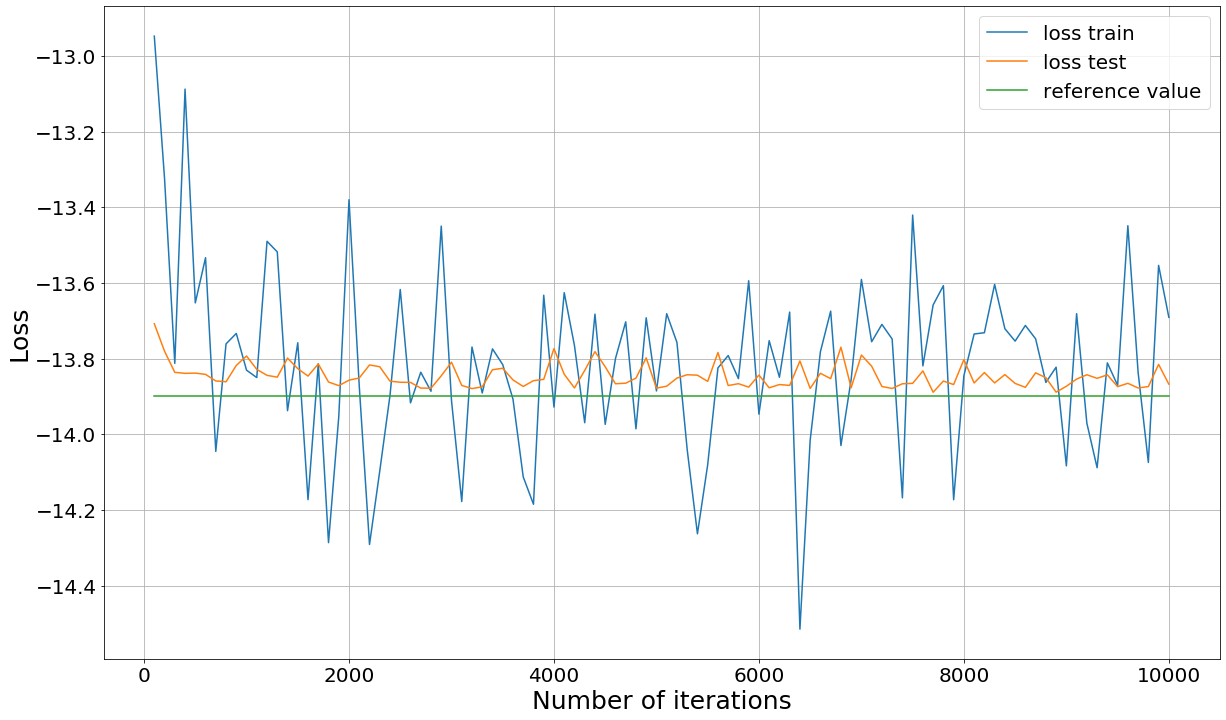}
        \caption{\it Max-call, $d=2$.}
    \end{subfigure}
        \begin{subfigure}[b]{0.45\textwidth}
        \centering
        \includegraphics[width=\textwidth]{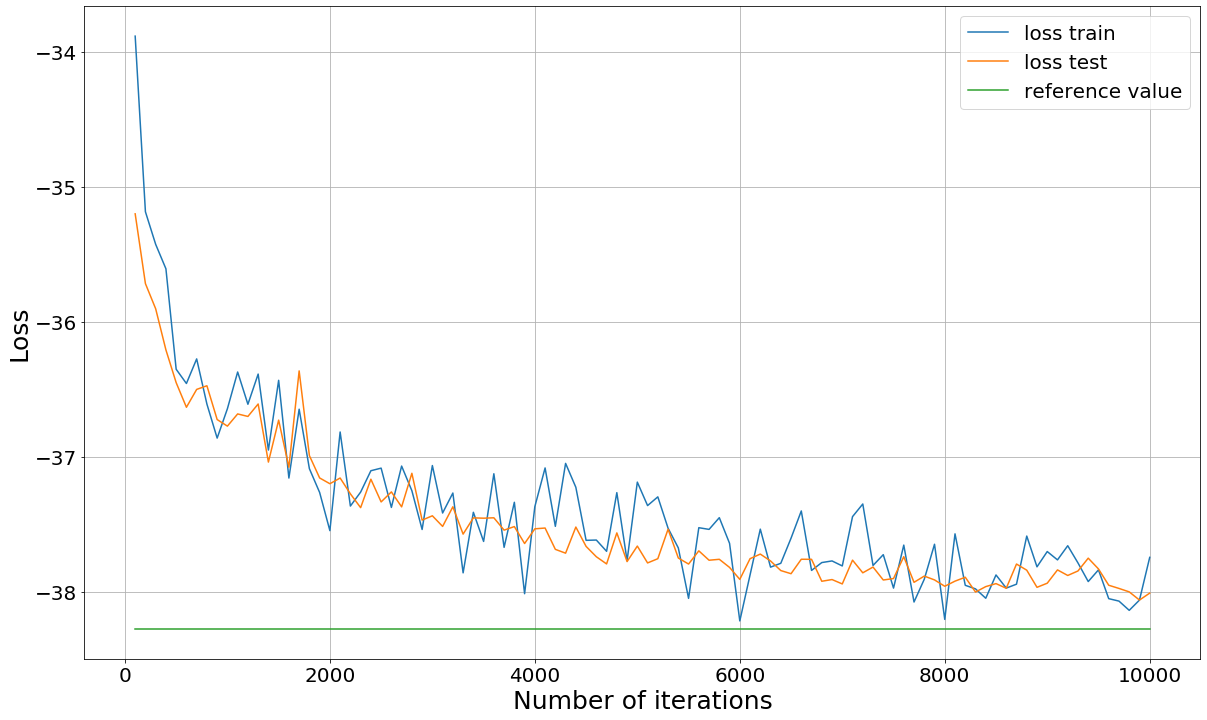}
        \caption{\it Max-call, $d=10$.}
    \end{subfigure}
            \begin{subfigure}[b]{0.45\textwidth}
        \centering
        \includegraphics[width=\textwidth]{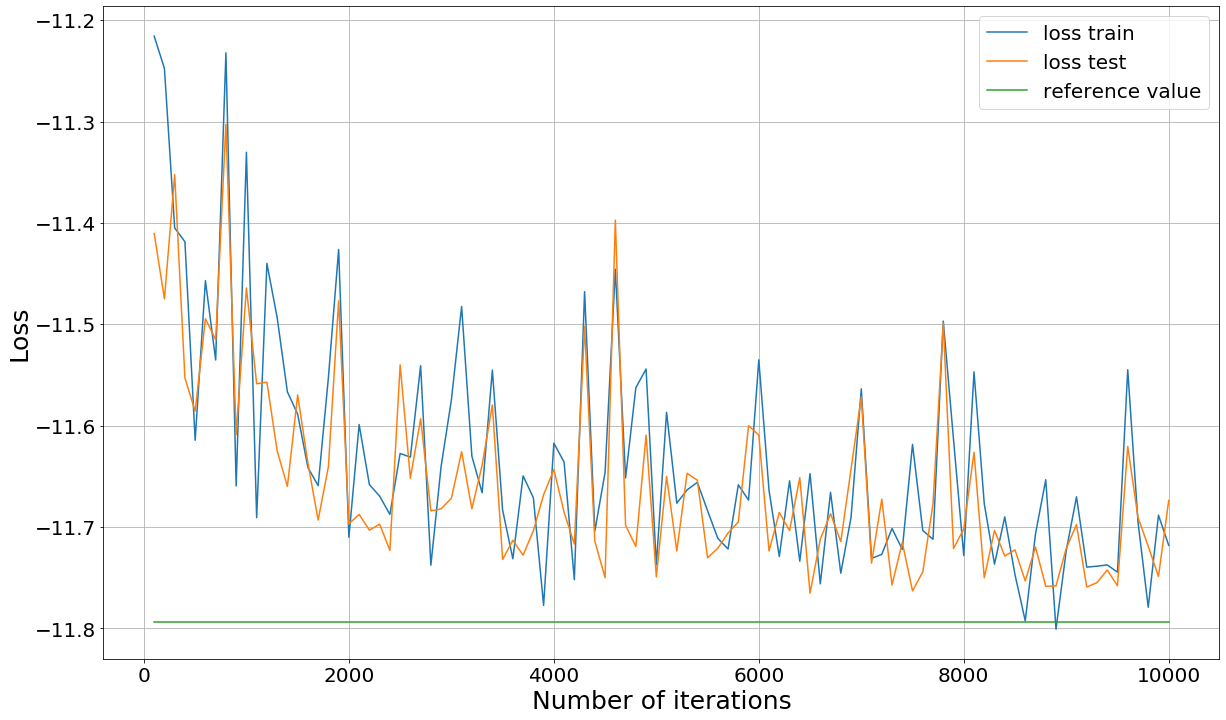}
        \caption{\it Strangle spread.}
    \end{subfigure}
     \caption{\it \label{fig:learningcurve}Learning curves for the different Bermudan options.}
\end{figure}

\bigskip
Losses and times obtained with Algorithm \ref{algo:algoGlobal} are given in Table \ref{tab:comparia} for each case and losses are compared to a reference value (\cite{bouchard08} for the put option and \cite{becker19b} for the other options). The algorithm succeeds in pricing Bermudan options with a high precision (relative error <1\%) in dimension up to 10 and number of time steps up to 50. The computing time is more sensitive to the number of time steps than to the dimension: the number of neural network estimation is equal to the number of time steps. The increase of computing time when dimension increases is mostly caused by a need to increase the batch size and a more important simulation time. Algorithm \ref{algo:algoGlobal} succeeds in pricing Bermudan options and solves problems that are usually hard to solve and very expensive in terms of computation time as they suffer from the curse of dimensionality. The training and testing learning curves are given in Figure \ref{fig:learningcurve}. The testing errors are relatively stable for the put and the max-call options but less stable for the strangle spread. For the put and the 2 dimensional max-call, the testing error converges quickly to the optimal value. The testing error of the 10 dimensional max-call decreases more slowly. The different training errors are all noisy as they are of smaller size.

\begin{figure}[!ht]
    \centering
    \begin{subfigure}[b]{\textwidth}
        \centering
        \includegraphics[width=\textwidth]{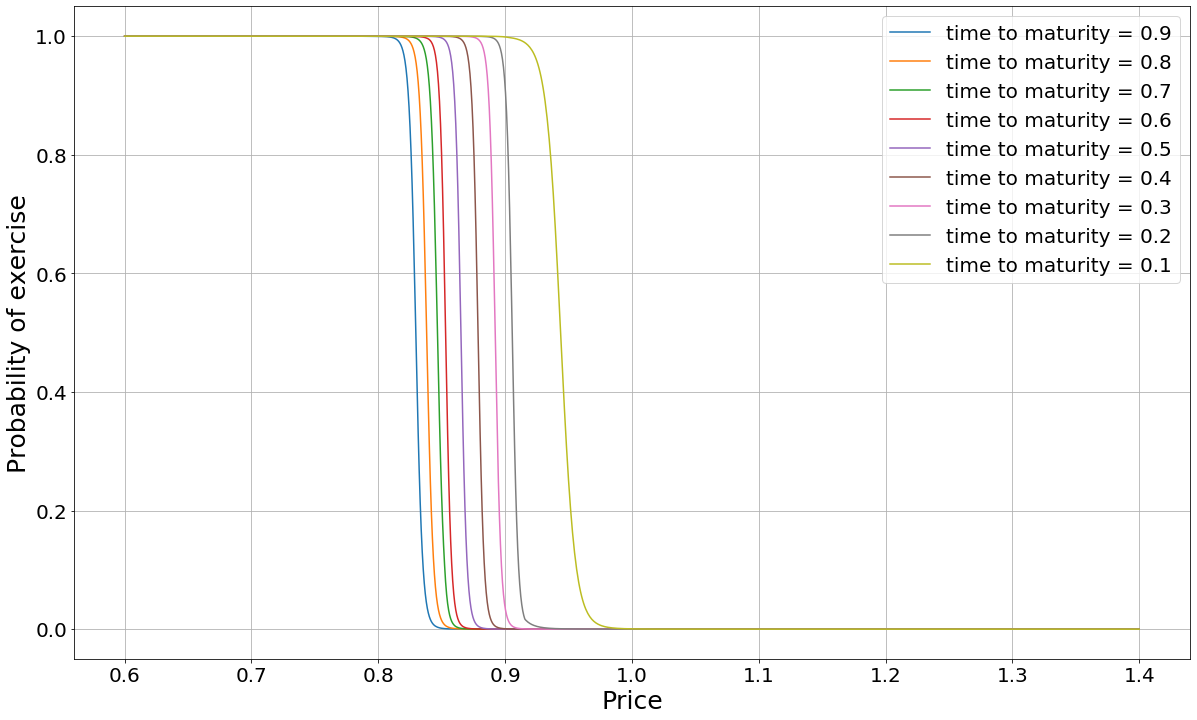}
        \caption{\it Probability of exercise.}
    \end{subfigure}
    \begin{subfigure}[b]{\textwidth}
        \centering
        \includegraphics[width=\textwidth]{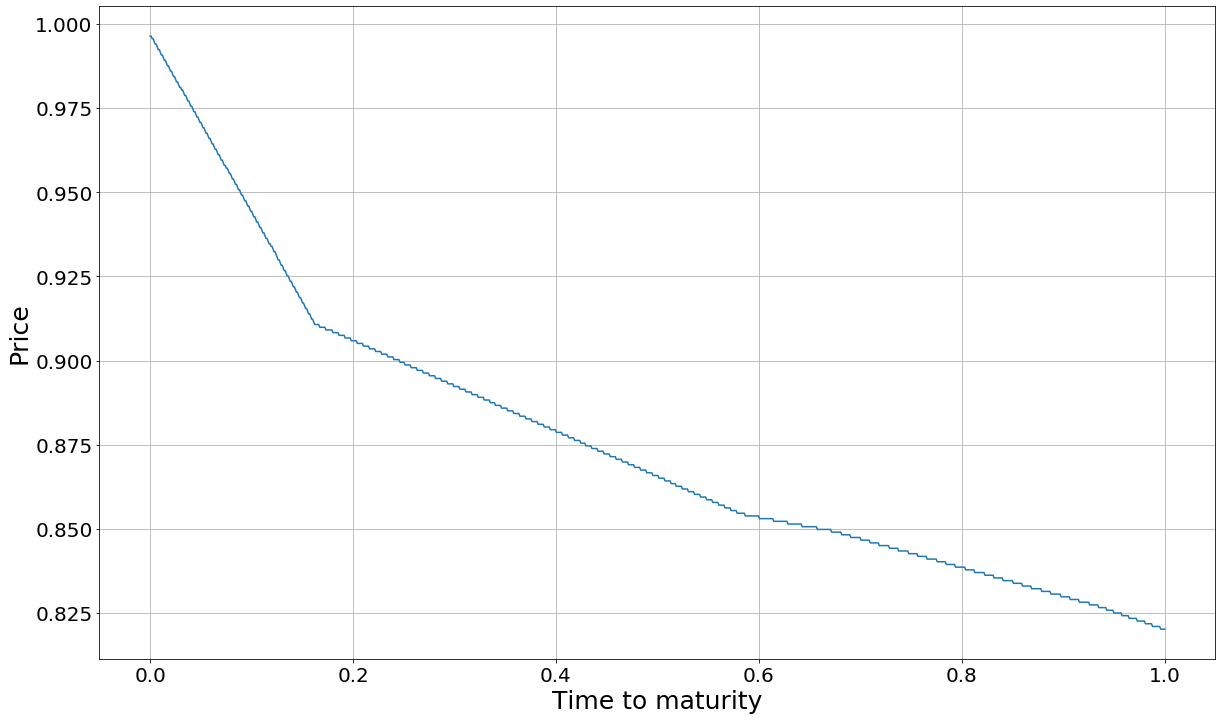}
        \caption{\it Exercise region.}
    \end{subfigure}
     \caption{\it \label{fig:bermudanex}Probability of exercise and exercise region (region below the curve) for the Bermudan put option computed from the trained neural network.}
\end{figure}

\bigskip
Once trained, the neural network allows to compute the probability to exercise according to the price and time to maturity and the exercise region in a few seconds, see Figure \ref{fig:bermudanex} for the Bermudan put option. The probability to exercise has a S-shape with limit values 0 and 1 (do not exercise and exercise) and a small transition region between those two values. As expected, the first value such that the probability becomes 0 increases when time to maturity decreases. This is confirmed in the exercise region in Figure \ref{fig:learningcurve} with the frontier decreasing with time to maturity. The exercise region is the one below the curve, which is computed using the first value such that the probability of exercise is below 0.5. The frontier is extrapolated from the neural network trained only on a discrete time grid but that allows to use different time to maturity values (even ones not used in the training), which is not possible with classical backward optimisation.

\begin{table}[!ht]
\centering
\begin{tabular}{lrrlr}
\toprule
Use case / Method &  Algorithm \ref{algo:algoGlobal} &  Reference  &  Difference &     Time (s) \\
\midrule
    Bermudan put &     0.0603 &     0.0603 &  0.06\% &    393.9 \\
  Max-call, d = 2 &    13.8787 &    13.8990 &  0.15\% &   1377.8 \\
 Max-call, d = 10 &    38.0347 &    38.2780 &  0.64\% &   5968.7 \\
  Strangle spread &    11.7681 &    11.7940 &  0.22\% &  12991.3 \\
\bottomrule
\end{tabular}
\caption{\it Results obtained on different Bermudan options pricing with Algorithm \ref{algo:algoGlobal} with the relative difference between Algorithm \ref{algo:algoGlobal} and a reference value (given by \cite{bouchard2012monte} for the put option and by \cite{becker19b} for the other options). The time in seconds corresponds to the time of training and predicting.}
\label{tab:comparia}
\end{table}

\subsection{Swing options without delay}
\label{sec:swing}

\begin{figure}[!ht]
    \centering
    \begin{subfigure}[b]{\textwidth}
        \centering
        \includegraphics[width=\textwidth]{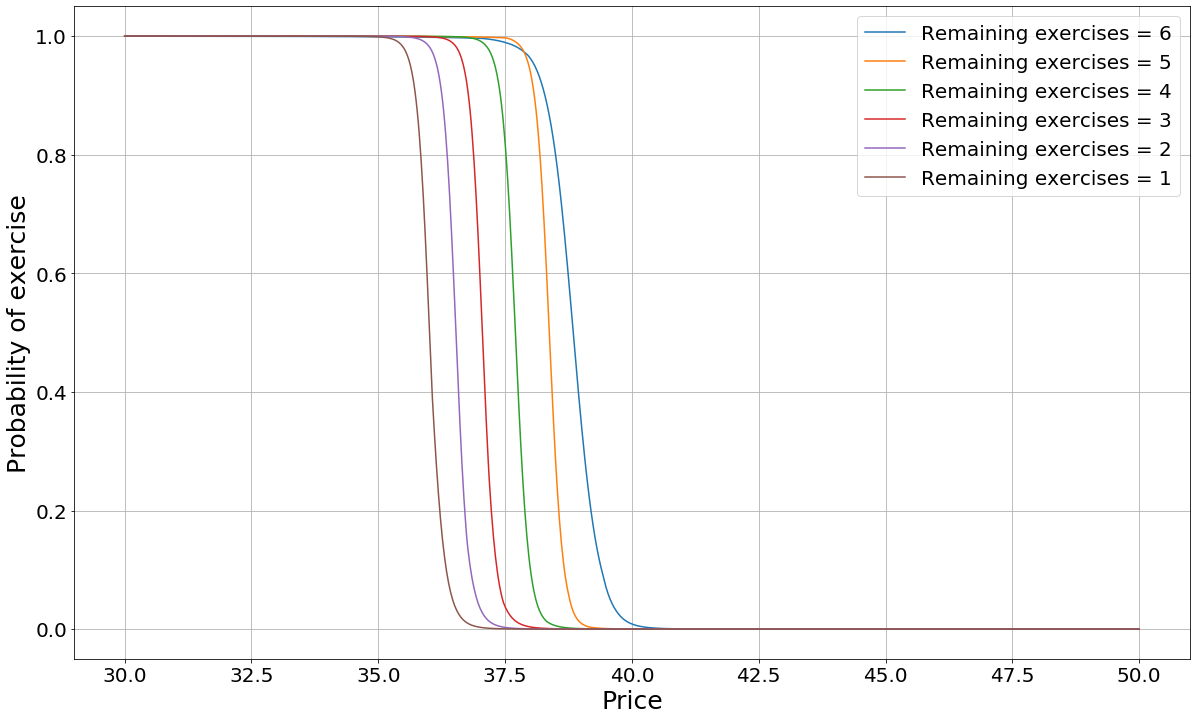}
        \caption{\it Probability of exercise at time $t=0.125$.}
    \end{subfigure}
    \begin{subfigure}[b]{\textwidth}
        \centering
        \includegraphics[width=\textwidth]{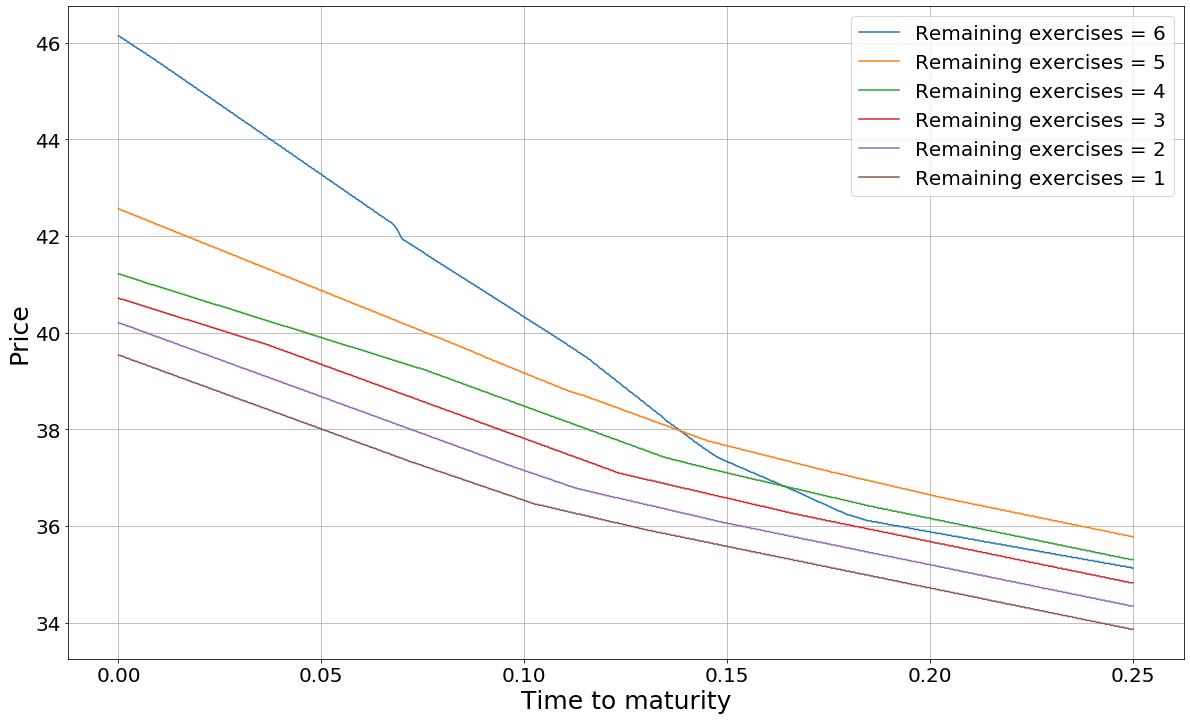}
        \caption{\it Exercise region.}
    \end{subfigure}
     \caption{\it \label{fig:swingex}Probability of exercise and exercise region (region below the curve) for the swing put option with strike 40 and maximum number of exercises 6.}
\end{figure}

In this section, we consider a swing option without delay constraint. We compare in Table \ref{tab:resultsswing} the results obtained by Algorithm \ref{algo:algoGlobal} with the results of \cite{ibanez04} in the case of a put option with $d=1$, $g(x) = (K-x)^+$, $K = 40$, $S_0 \in \{35, 40, 45\}$, $r = 0.0488$, $\delta = 0$, $\Sigma = 0.25$, $N = 12$, $T=0.25$, $l \in \{1, 2, 3, 4, 5, 6\}$ and no delay. We consider a batch size equal to $N_{batch} = 2,000$, a neural networks with 3 layers size 10 and $N_{iter} = 5,000$ iterations. Every case takes around 4 minutes to converge, see Table \ref{tab:resultsswingtimes}. The algorithm gives very accurate results in a short period of time for the valuation of the swing options. As for the Bermudan put option, it is possible to compute the probability of exercise and the exercise region for this swing option, see Figure \ref{fig:swingex}. Those quantities depend now on the remaining number of exercises. As expected, the first value such that the probability to exercise is 0 increases with the remaining number of exercises. The exercise frontier should increase with the remaining number of exercises which is the case most of the time : the curve for a remaining number of exercises equal to 6 goes below the ones with a remaining number of exercises equal to 5 (resp. 4) when time to maturity is greater than 0.1 (resp. 0.15).

\begin{table}[ht!]
\centering
\begin{tabular}{rlll}
\toprule
 $l$ / $S_0$ &                        35 &                      40 &                      45 \\
\midrule
           1 &    (5.104, 5.114, 0.19\%) &  (1.776, 1.774, 0.12\%) &  (0.409, 0.411, 0.42\%) \\
           2 &  (10.165, 10.195, 0.29\%) &   (3.492, 3.48, 0.34\%) &  (0.772, 0.772, 0.06\%) \\
           3 &   (15.194, 15.23, 0.24\%) &  (5.115, 5.111, 0.07\%) &   (1.09, 1.089, 0.09\%) \\
           4 &   (20.188, 20.23, 0.21\%) &  (6.658, 6.661, 0.05\%) &   (1.358, 1.358, 0.0\%) \\
           5 &     (25.19, 25.2, 0.04\%) &   (8.148, 8.124, 0.3\%) &   (1.58, 1.582, 0.12\%) \\
           6 &  (30.156, 30.121, 0.12\%) &  (9.494, 9.502, 0.09\%) &  (1.764, 1.756, 0.44\%) \\
\bottomrule
\end{tabular}
\caption{\it Comparison of results obtained by Algorithm \ref{algo:algoGlobal} with the ones of \cite{ibanez04} for different initial values $S_0$ and different number of executions $l$. The first value corresponds to the swing option value obtained with Algorithm \ref{algo:algoGlobal}, the second value to the one in \cite{ibanez04} and the third value is the relative difference in $\%$.}
\label{tab:resultsswing}
\end{table}

\begin{table}[ht!]
\centering
\begin{tabular}{rrrr}
\toprule
 $l$ / $S_0$ &     35 &     40 &     45 \\
\midrule
                                   1 &  270.1 &  278.6 &  249.9 \\
                                   2 &  241.6 &  246.6 &  236.8 \\
                                   3 &  240.6 &  237.7 &  243.4 \\
                                   4 &  239.6 &  242.4 &  240.0 \\
                                   5 &  271.3 &  239.9 &  235.3 \\
                                   6 &  243.8 &  240.9 &  241.7 \\
\bottomrule
\end{tabular}
\caption{\it Time in seconds for training and predicting with Algorithm \ref{algo:algoGlobal} to price the swing put option for different initial values $S_0$ and different number of executions $l$. }
\label{tab:resultsswingtimes}

\end{table}

\medskip
To assess the performance of Algorithm \ref{algo:algoGlobal} in high dimension, let us consider the pricing of the geometrical put option having payoff $g(x) = (K - \prod_{i=1}^d x_i)^+$. Let $d = 5$, $K = 40$, $S_0^i = 40^{1/5}$, $r = 0.0488$, $\delta = \frac{4r}{5}$, $\Sigma = \frac{0.25}{\sqrt{5}} I_5$, $N = 12$, $T=0.25$ and $l \in \{1, 2, 3, 4, 5, 6\}$. Prices dynamic parameters are chosen in order to have an option value equal to the one dimensional case put option value: the product of the components of $X$ follows a Black-Scholes dynamic with drift parameter equal to $0.0488$ and volatility equal to $0.25$. It allows to have a reference value (from \cite{ibanez04}) while considering a high dimensional case. We consider a batch size equal to $N_{batch} = 8,000$, a neural networks with 3 layers of size 30 and $N_{iter} = 5,000$ iterations. Results are given in Table \ref{tab:swing5d}. The algorithm succeeds in pricing this option with 5 underlyings in a reasonable time (less than 30 minutes). By using a less costly hyperparameterization (20 neurons density instead of 30, $2,000$ iterations instead of $5,000$, $N_{batch} = 3,000$ instead of $8,000$ we are able to obtain results in less than 4 minutes with a $2\%$ accuracy as shown in Table \ref{tab:swing5dlight}

\begin{table}[ht!]
\centering
\begin{tabular}{lrrlr}
\toprule
Use case / Method &  Algorithm \ref{algo:algoGlobal} &  Reference  &  Difference &    Time (s)\\
\midrule
            l = 1 &      1.767 &      1.774 &  0.39\% &  1814.3 \\
            l = 2 &      3.478 &      3.480 &  0.04\% &  1686.6 \\
            l = 3 &      5.100 &      5.111 &  0.22\% &  1701.5 \\
            l = 4 &      6.639 &      6.661 &  0.32\% &  1661.0 \\
            l = 5 &      8.117 &      8.124 &  0.09\% &  1689.2 \\
            l = 6 &      9.478 &      9.502 &  0.25\% &  1675.7 \\
\bottomrule
\end{tabular}
\begin{tabular}{lrrlr}
\end{tabular}
\caption{\it Comparison of results obtained by Algorithm \ref{algo:algoGlobal} for the pricing of a 5 dimensional swing put option with the reference values reported in \cite{ibanez04}. The time in seconds corresponds to the time of training and predicting.}
\label{tab:swing5d}
\end{table}

\begin{table}[ht!]
\centering
\begin{tabular}{lrrlr}
\toprule
Use case / Method &  Algorithm \ref{algo:algoGlobal} &  Reference & Difference &  Time (s) \\
\midrule
            l = 1 &                            1.733 &      1.774 &      2.33\% &     221.9 \\
            l = 2 &                            3.435 &      3.480 &      1.29\% &     222.0 \\
            l = 3 &                            5.074 &      5.111 &      0.71\% &     212.2 \\
            l = 4 &                            6.591 &      6.661 &      1.05\% &     199.4 \\
            l = 5 &                            8.033 &      8.124 &      1.13\% &     197.0 \\
            l = 6 &                            9.392 &      9.502 &      1.16\% &     203.0 \\
\bottomrule
\end{tabular}

\caption{\it Comparison of results obtained by Algorithm \ref{algo:algoGlobal}  with suboptimal hyperparameters for the pricing of a 5 dimensional swing put option with the reference values reported in \cite{ibanez04}. The time in seconds corresponds to the time of training and predicting.}
\label{tab:swing5dlight}
\end{table}

\subsection{Swing options with delay}
\medskip
Let us now consider the case of a put option with $d=1$, $g(x) = (K-x)^+$, $K = 100$, $S_0 = 100$, $r = 0.05$, $\delta = 0$, $\Sigma = 0.3$, $N = 50$, $T=1$, $\gamma = 5 \frac{T}{N}$ and $l \in \{1, 2, 3, 4, 5\}$. Delay constraint is now present and a higher number of dates is considered. We consider a batch size equal to $N_{batch} = 5,000$, a neural networks with 3 layers of size 10 and $N_{iter} = 10,000$ iterations. We compare in Table \ref{tab:resultsswing2} the results obtained with Algorithm \ref{algo:algoGlobal} to the ones obtained with \cite{carmona08}. The algorithm gives satisfying results but in this situation we can see that the relative error increases with the number of exercises. 

\begin{table}[ht!]
\centering
\begin{tabular}{lrrlr}
\toprule
Use case / Method &  Algorithm \ref{algo:algoGlobal} &  Reference &  Difference &    Time (s) \\
\midrule
            l = 1 &                            9.844 &       9.85 &     0.06\% &    5430.5 \\
            l = 2 &                           19.093 &      19.26 &     0.87\% &    6658.0 \\
            l = 3 &                           27.827 &      28.80 &     3.38\% &    5518.4 \\
            l = 4 &                           36.058 &      38.48 &     6.29\% &    5566.1 \\
            l = 5 &                           43.638 &      48.32 &     9.69\% &    5472.1 \\
\bottomrule
\end{tabular}
\caption{\it Comparison of results obtained by Algorithm \ref{algo:algoGlobal} with the reference values reported in \cite{carmona08}. The time in seconds corresponds to the time of training and predicting.}
\label{tab:resultsswing2}
\end{table}

\bigskip

%% file: 3_conclusion.tex
\section{Conclusion and perspectives}
A stochastic control algorithm able to deal with (discrete) optimal stopping variables is presented. The different use cases show that the proposed algorithm is able to solve optimal stopping time problems in a reasonable time, even when the dimension is high and also for multi-exercise. The algorithm is simple and allows us to find an optimal policy without any knowledge on the dynamic programming equation. The method presented in this paper avoids a costly backward pass and only need a forward pass. While the computation time increases a little with dimension, it increases a lot more with the number of time steps and the algorithm can have troubles to converge. To confirm all those results, one should study the theoretical convergence of the algorithm. This algorithm could easily be extended to impulse control if combined with \cite{fecamp2019risk} in order to solve problems involving both continuous and discrete controls such as hedging with fixed transaction costs.